# Simple and accurate method to simulate resistors and wires in a nanoscale circuit

Mark J. Hagmann and Logan D. Gibb
NewPath Research LLC., 2880 S. Main Street, Suite 214, Salt Lake City, Utah, USA 84115
(Dated May 9, 2020)

In solving the Schrödinger equation to simulate a nanoscale circuit, we note that the mean free path for electrons in some metals is as large as 48 nm. Thus, the wavefunction may propagate coherently through wires corresponding to the lines that show the potential outside of the tunneling junction. A voltage source may be modeled as a jump in the potential. Similarly, the potential across a resistor may be modeled as a sharp drop or a downward sloping line to show a decrease in the potential. Then the resistance may be determined by dividing this voltage drop by the product of the calculated current density and the effective cross-sectional area.

## I. INTRODUCTION

Previously we applied the transfer-matrix method to simulate the effects of the finite transit time in laser-assisted quantum tunneling [1] and discovered a resonance caused by the transit time [2]. These numerical simulations guided us in developing microwave oscillators based on laser-assisted field emission [3] and generating microwave frequency combs by focusing a mode-locked laser on the tunneling junction of a scanning tunneling microscope (STM) [4]. Currently we are studying a variant of the STM where extremely low-power ($\approx$ 3 atto-watt) microwave harmonics of the laser pulse repetition rate have a signal-to-noise ratio of 20 dB [5]. These extremely low-noise room-temperature measurements are possible because the quality factor (Q) is approximately $10^{12}$, which is 5 times the Q of a cryogenic microwave cavity [6]. Now we describe new methods for the analytical procedures that we are developing to further guide this effort.

## II. SIGNIFICANCE OF COHERENT TRANSPORT IN A NANOSCALE CIRCUIT

In solving the Schrödinger equation to simulate a nanoscale circuit, we note that the mean free path for electrons in some metals is as large as 48 nm [7]. Thus, the wavefunction may propagate coherently through wires that correspond to the lines showing the potential outside the tunneling junction. We acknowledge that, surprisingly, the effective resistance of a nanoscale wire is proportional to the mean free path [7]. This effect will be studied in greater detail before devices are made [8],[9]. We are particularly interested in nanoscale devices that are based on quantum tunneling by (1) Field emission where the transport of electrons between two electrodes is partly by tunneling and partly classical [3], and (2) Direct tunneling between two electrodes [4]. Both field emission and direct tunneling introduce high values of electrical resistance so we anticipate that the increased resistance of the wires may not present a serious problem.

Previously others have neglected the full interaction within a nanoscale circuit. For example, they have assumed an incident and reflected wave at one end of the barrier and a transmitted wave at the other end, without considering how these three waves may interact at the voltage source [10]. Others have modeled tunneling junctions by assuming that the anode and cathode are held at fixed potentials without considering the complete circuit [11],[12]. At the present time we are studying the use of a consistent closed-loop nanoscale model as shown in Figure 1 to allow for the full effects in the circuit due to the large mean free path. Note that the source of dc potential is between the terminal on the RHS and the ground at the LHS. The symbols

U and V are used for the potential energy and the voltage to avoid confusion. The use of a sloped line to represent the load resistor is explained in the following section of this paper.

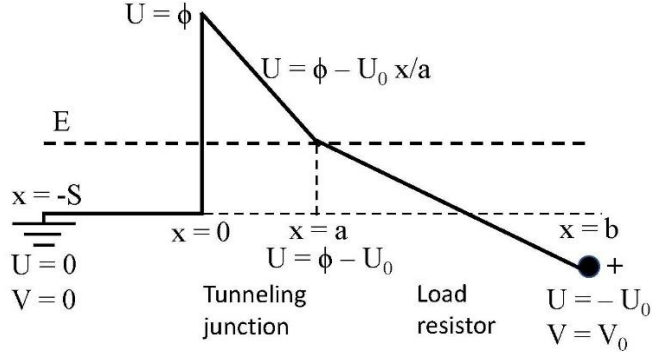

FIG. 1: Model for a closed-loop solution using a tunneling junction with a load resistor.

### III. MODELS FOR RESISTORS AND CONNECTING WIRES

In quantum simulations a voltage source is generally modeled as a jump in the potential at a specific location but it could be represented by a linear rise in the potential over a specified distance. In Fig. 1 note that the voltage source is between the point at the RHS where the potential is $V_0$ and the ground at the LHS of the model. In this figure the load resistor is modeled by a linear fall in the potential over a specified distance.

Figure 2 shows a circuit consisting of only a load resistor, a voltage source, a lossless wire connecting them and two short lossless wires connected to a common ground.

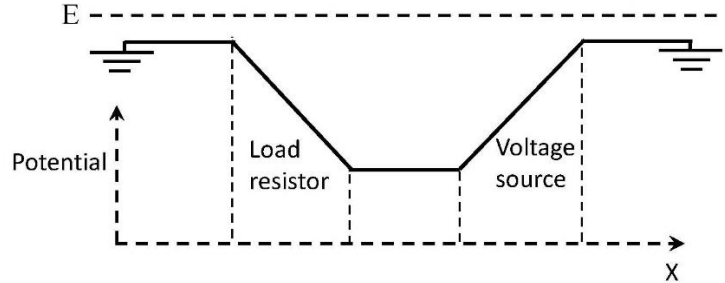

FIG. 2: Symmetric model consisting of a load resistor, a voltage source, and lossless connectors.

For a static problem the current density in the x-direction is given by Eq. (1), as the product of the probability current density and the electron charge. Thus, the effective value of the resistance may be determined by dividing the voltage drop across the simulated load resistor by the product of the electrical current density and the effective cross-sectional area.

$$J_X(x) = \frac{-ie\hbar}{2m}\left(\psi \frac{d\psi^*}{dx} - \psi^* \frac{d\psi}{dx}\right) \qquad (1)$$

It would also be possible to model a resistor as a sharp drop in the potential as we model a voltage source by a jump in the potential. It is essential for the potential to be below the energy E at each point in the load resistor to avoid quantum tunneling. Adding resistors to a model is the converse to adding voltage sources. Horizontal lines, with constant potential, represent wires having no significant resistivity.

Analytical solutions for the wavefunction with a linear fall in the potential to represent a resistor are possible using Airy functions [13]. In an application the voltage source could be a nanoscale capacitor fed by lossy leads to an external power supply.

## IV. QUASISTATIC APPROXIMATION FOR TIME-DEPENDENT POTENTIALS

It is possible to have two or more jumps in the potential to simulate separate voltage sources. For example, one may be an applied DC bias and the other may be sinusoidal or have other time-dependence.

In 1963 Tien and Gordon [14] published the first analytical solution of the time-dependent Schrödinger equation for electron tunneling in a sinusoidally-modulated triangular potential barrier. Google Scholar lists 994 papers that refer to this analysis when studying quantum tunneling in a variety of nanostructures. Tien and Gordon considered the effect of superimposing a sinusoidal voltage on the dc voltage at the anode of a vacuum tunneling diode where the cathode is grounded. Their solution shows that adding the sinusoidal voltage does not change the spatial dependence of the wavefunction but only modifies the electron energies by adding and subtracting quanta at integer multiples of the frequency of the sinusoid. Thus, adding the sinusoidal voltage causes no change in the current with metal electrodes but does change the measured current for the special case of superconductor electrodes. We acknowledge that this is a valid solution but do not regard it as being unique, and we question many of the applications by others which do not pertain to superconductors.

Consider what would happen if the DC potential was adjusted up and down by "turning a knob" on a DC power supply connected to a tunneling junction. This would cause the current to alternately increase and decrease—but how fast can this change be made with a proportional response? The answer is that this quasistatic approximation is only appropriate at frequencies that are below that where photon processes must be considered. Consider the interaction of the radiation from a laser with a tunneling junction. For example, with a laser having a wavelength of 700 nm a tunneling junction having a length of 0.5 nm is only 0.07 percent of the wavelength so this would be considered to be quasistatic.

An analysis of the data from our measurements of the microwave frequency comb generated by focusing a mode-locked laser on the tunneling junction of an STM shows that our data are consistent with a quasistatic approximation [15]. Thus, when we have completed the solution for static models of nanoscale circuits, we will use quasistatic approximations to extend this method to guide our work in developing new devices for microwave and terahertz frequencies.

## V. CONCLUSIONS

It appears that this is the first time that quantum effects have been modeled consistently throughout an entire static nanoscale circuit, including the modeling of resistors and lossy wires. An analysis of the data from our experiments with laser-assisted STM suggests that quasistatic approximations are adequate. Thus, when our static analysis is complete, we will use quasistatic approximations to extend this model to laser-assisted tunneling in nanoscale circuits.

## Acknowledgments

We are grateful to Dmitry Yarotski who made it possible for the first author to make measurements in laser-assisted Scanning Tunneling Microscopy in visits to the Center for Integrated Nanotechnologies (CINT) at Los Alamos National Laboratory from 2008 to 2017. This work was sponsored by the National Science Foundation under Grant 1648811 and the U.S. Department of Energy under Award DE-SC000639.